\documentclass[11pt]{article}
\pdfoutput=1 
\usepackage{a4wide}
\usepackage{epsfig,epsf,graphics}
\usepackage{times}
\usepackage{color}
\usepackage{amsfonts,amssymb,latexsym,amsmath} 
\usepackage{bm}
\usepackage{enumerate}
\usepackage{enumitem}
\usepackage{hhline}
\usepackage{longtable}
\usepackage{array} 
\usepackage[
colorlinks=true,
linkcolor=black,
breaklinks=true,
urlcolor=green,
citecolor=blue]{hyperref}
\usepackage{cite}
\usepackage{epstopdf}

\newcommand{\al}{&\!\!\!\!}
\newcommand{\Lag}{\mathcal{L}}

\newcommand{\Tr}{\textrm{Tr}}

\newcommand{\X}{X_2(4012)}
\newcommand{\Y}{Y(4260)}

\newcommand{\email}[1]{\footnote{{\em E-mail address:} \texttt{#1}}}

\begin{document}
\title{Production of the spin partner of the $X(3872)$ in $e^+e^-$ collisions}

\author{Feng-Kun Guo$^{\, a,}$\email{fkguo@hiskp.uni-bonn.de} ,
Ulf-G.~Mei{\ss}ner$^{\, a,b,}$\email{meissner@hiskp.uni-bonn.de} ,
Zhi Yang$^{\, a,}$\email{zhiyang@hiskp.uni-bonn.de}\\ 
   {\it\small$^a$Helmholtz-Institut f\"ur Strahlen- und Kernphysik and Bethe
   Center for Theoretical Physics, }\\
   {\it\small Universit\"at Bonn,  D-53115 Bonn, Germany }\\
   {\it\small$^b$Institut f\"{u}r Kernphysik, Institute for Advanced
Simulation, and J\"ulich Center for Hadron
Physics,}\\
   {\it \small Forschungszentrum J\"ulich, D-52425 J\"{u}lich, Germany}
   }

\maketitle

\begin{abstract}
\noindent
We study the production of the spin partner of the $X(3872)$,
which is a $D^{*}\bar D^{*}$ bound state with quantum numbers $J^{PC}=2^{++}$
and named $X_2(4012)$ here, with the associated  emission
of a photon in electron--positron collisions.
The results show that the ideal energy region to observe the $\X$ in
$e^+e^-$ annihilations is from 4.4~GeV to 4.5~GeV, due to the
presence of the $S$-wave $\bar D^{*} D_1(2420)$ and $\bar D^{*} D_2(2460)$
thresholds, respectively. We also point out that
it will be difficult to observe the  $\gamma X(4012)$ at the $e^+e^-$
center-of-mass energy around 4.26~GeV.

\end{abstract}

\newpage

In the heavy quarkonium mass region, the so called $XYZ$ states have been
observed, and many of these quarkonium-like states defy a conventional quark
model interpretation.
They are therefore suggested to be exotic.
The $X(3872)$, discovered by the Belle Collaboration~\cite{Choi:2003ue}, is the
one of the most interesting exotic states. As the mass of the $X(3872)$ is
extremely close to the $D^{0}\bar D^{*0}$ threshold, it is regarded as one
especially promising candidate for a hadronic molecule.

Effective field theory (EFT) can cope with the interaction between  heavy mesons
in bound state systems at low energies.
For such a kind of systems,  heavy quark symmetry is relevant due to the
presence of the heavy quark/antiquark in the meson/antimeson. This fact leads to
predictions of new states as partners of the observed $XYZ$ states in the hadron
spectrum.
For example, with an EFT description of the heavy mesonic molecules, the heavy
quark symmetry can be used to predict the existence of the spin and bottom
partners of the $X(3872)$~\cite{Nieves:2012tt,Guo:2013sya}.

The spin partner of the $X(3872)$, called $\X$ hereafter, is predicted to
exist as the $S$-wave bound state of $D^{*}\bar D^{*}$ with quantum numbers
$2^{++}$~\cite{Nieves:2012tt}. Such a state was also expected to exist in other
models, see
Refs.~\cite{Tornqvist:1993ng,Wong:2003xk,Swanson:2005tn,Molina:2009ct,Sun:2012zzd}.
It is different from the $X(3872)$ in several aspects: first, being an isoscalar
state it should decay into the $J/\psi\pi\pi\pi$ with a branching fraction much
larger than that for the $J/\psi\pi\pi$ because the $J/\psi\rho$ and $J/\psi\omega$ thresholds are far
below the mass of the $\X$ (very different to the case of the $X(3872)$); second,
it is expected to decay dominantly into open charm mesons, $D\bar D$ , $D\bar
D^*$ and $D^*\bar D$, in a $D$-wave with a width of the order of a few
MeV~\cite{Guo:unpublished}; third, its mass as set by the $D^*\bar D^*$
threshold is higher than the quark model prediction for the first
radially excited $\chi_{c2}$~\cite{Godfrey:1985xj}.

The significance of the $X_2(4012)$ state is that its mass should be
approximately given by the
\begin{equation}
  M_{X_2(4012)} \approx M_{X(3872)} + M_{D^*} - M_{D} \approx 4012~\text{MeV}
\end{equation}
as dictated by heavy quark spin symmetry for heavy-flavor hadronic
molecules~\cite{Guo:2009id,Guo:2013sya}. Notice that a state with
the same quantum numbers $2^{++}$ was also predicted in the tetraquark
model~\cite{Maiani:2004vq}. However, the fine splitting between the $2^{++}$ and
$1^{++}$ tetraquarks, which was predicted to be 70~MeV in
Ref.~\cite{Maiani:2004vq}, is not locked to that between the $D^*$ and $D$.
Similarly, the splitting between the $2P$ $c\bar c$ states in the Godfrey--Isgur
quark model is 30~MeV~\cite{Godfrey:1985xj}, also much smaller than
$M_{D^*}-M_D$.
Therefore, if a $2^{++}$ state will be observed in experiments with a mass
around 4012~MeV, the mass by itself would already be a strong support for the
hadronic molecular nature of both the $X(3872)$ and the tensor state. As a
result, searching for a $2^{++}$ state with a mass around 4012~MeV
is very important even for understanding the nature of the $X(3872)$.

However, although the $X(3872)$ has been
observed by many other experiments after its discovery~\cite{Aubert:2004ns,Acosta:2003zx,Abazov:2004kp,Aaij:2011sn,Aaij:2013zoa,Chatrchyan:2013cld},
no evidence for the existence of its spin partner has been reported.
In Ref.~~\cite{Guo:2014sca}, it is shown that the prompt production of the $\X$
presents a significant discovery potential at hadron colliders. In this paper,
we will investigate the production of the $\X$ associated with the photon
radiation in electron--positron collisions. This work presents an extension of
the study on the production of the $X(3872)$ as a $D\bar D^*$ molecule in
charmonia radiative transitions reported in Ref.~\cite{Guo:2013zbw}.
In that paper, it was shown that the
favorite energy regions for the $X(3872)\gamma$ production are around the
$Y(4260)$ mass and 4.45~GeV. Later on, the BESIII Collaboration observed events
for the process $Y(4260)\to X(3872)\gamma$~\cite{Ablikim:2013dyn}, which may be
regarded as a support of the dominantly molecular nature of the $X(3872)$. Since
the existence of the $D^*\bar D^*$ bound state, the $\X$, is the consequence of
the heavy quark spin symmetry of the molecular nature of the $X(3872)$,
the production of the $\X$ in $e^+e^-$ collisions in the energy range of the
BESIII experiment~\cite{Asner:2008nq} thus provides an opportunity to search
for new charmonium-like states on the one hand and can offer useful information
towards understanding the $X(3872)$ on the other hand.

\begin{figure}[tb]
\begin{center}
  \includegraphics[width=0.8\textwidth]{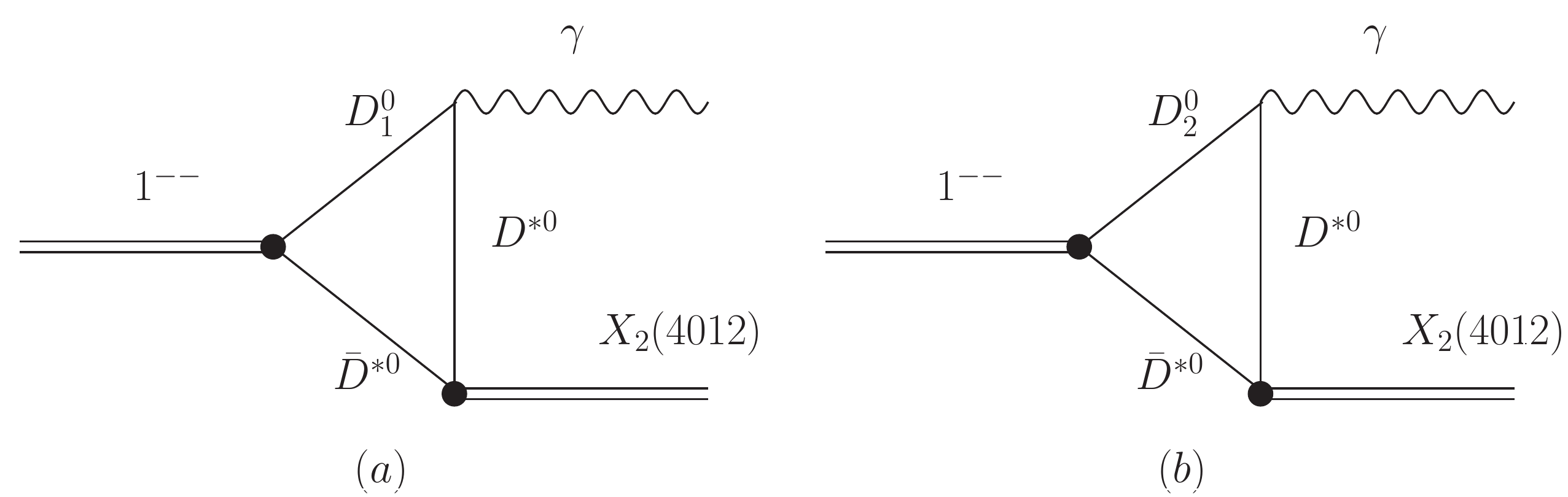}
\caption{Relevant triangle diagrams for the production of the $\X$ in
vector charmonia radiative decays. The charge-conjugated diagrams are not shown here.
}
\label{fig:FeynmanDiagram}
\end{center}
\end{figure}

The production of the $X(3872)$ through the radiative decay of the $\psi(4160)$
charmonium is considered in Ref.~\cite{Margaryan:2013tta} using  heavy hadron
chiral perturbation theory along with the X-EFT~\cite{Fleming:2007rp}. Then,
Ref.~\cite{Guo:2013zbw} studied the $X(3872)$ production by considering the
contribution from intermediate charmed meson loops, and it was argued that the
dominant mechanism is as follows: the initial charmonium is coupled to a pair of
charmed mesons with one being $S$-wave with $s_\ell^P=\frac12^-$, where $s_\ell$
is the total angular momentum of the light-flavor cloud in the charmed meson,
and the other being $P$-wave with $s_\ell^P=\frac32^+$, and the $P$-wave charmed
meson radiatively transits to a $D (D^*)$ which coalesce with the other $S$-wave
charmed meson, $\bar D^* (\bar D)$, into the $X(3872)$. The spin partner of
$X(3872)$, the $\X$, can be produced by a similar mechanism as shown in
Fig.~\ref{fig:FeynmanDiagram}.
Notice that the $\X$ couples to $D^*\bar
D^*$ instead of $D\bar D^*+c.c.$, as it is in the case of the $X(3872)$. We will
only consider the neutral charmed mesons in the loops because the photonic
coupling between the $P$-wave and $S$-wave charmed mesons for the neutral ones
is much larger than that for the charged. This is due to cancellation of
contributions from the charm and down quarks in the charged mesons, see,
e.g.~\cite{Fayyazuddin:1993cc}.
In the loops, the $\X$ couples to the $D^{*0} \bar D^{*0}$ pair in an $S$-wave.
With the quantum numbers being $1^{--}$, the initial charmonium can couple to
one $P$-wave and one $S$-wave charmed meson in either $S$- or $D$-wave. Since
both the initial charmonium and the $\X$ in the final state are close to the
corresponding thresholds of the charmed-meson pairs, we are able to use a power
counting in velocity of the intermediate mesons. Following the power counting
rules as detailed in Ref.~\cite{Guo:2010ak} and presented in the case of
interest in Refs.~\cite{Guo:2013zbw,Guo:2014qra}, the dominant contribution
comes from the case when the coupling of the initial charmonium to the charmed
mesons is in an $S$-wave. In this case, the initial charmonium should be a
$D$-wave state in the heavy quark limit $m_c\to\infty$ as a consequence of
heavy quark spin symmetry~\cite{Li:2013yka}.

The charmed mesons can be classified according to the total angular momentum of
the light degrees of freedom $s_\ell$ and collected in doublets with total spin
$J=s_\ell\pm \frac{1}{2}$ in the heavy quark limit.
The $s_\ell^P=\frac{1}{2}^{-}$ states correspond to charmed mesons in the
doublet $(0^{-},1^{-})$, here denoted as $(P,V)$, whereas the
$s_\ell^P=\frac{3}{2}^{+}$ states correspond to charmed mesons in the doublet
$(1^{+},2^{+})$, denoted as $(P_1,P_2)$.
To describe these heavy mesons, we choose the two-component notation introduced
in Ref.~\cite{Hu:2005gf}. The notation uses $2\times 2$ matrix fields, and is
convenient for nonrelativistic calculations.
The fields for the relevant heavy meson states are
\begin{eqnarray}
  H_a \al=\al \vec{V}_a\cdot \vec{\sigma}+ P_a, \nonumber\\
  T_a^i \al=\al P_{2a}^{ij} \sigma^j + \sqrt{\frac23}\, P_{1a}^i + i \sqrt{\frac16}\,
\epsilon_{ijk} P_{1a}^j \sigma^k,
  \label{eq:hta}
\end{eqnarray}
for the $s_\ell^P=\frac{1}{2}^{-}$ ($S$-wave) and $s_\ell^P=\frac{3}{2}^+$
($P$-wave) heavy mesons, respectively, where $\vec\sigma$ are the Pauli
matrices, and $a$ is the flavor index for the light quarks. In
Eq.~\eqref{eq:hta}, $P_a$ and $V_a$ annihilate the pseudoscalar and vector heavy
mesons, respectively, and $P_{1a}$ and $P_{2a}$ annihilate the excited
axial-vector and tensor heavy mesons, respectively. Under the same phase
convention for charge conjugation specified in Ref.~\cite{Guo:2013zbw},
the fields annihilating the mesons containing an anticharm quark
are~\cite{Fleming:2008yn}
\begin{eqnarray}
  \bar H_a \al=\al -\vec{\bar V}_a\cdot \vec{\sigma} + \bar P_a, \nonumber\\
  \bar T_a^i \al=\al -\bar P_{2a}^{ij} \sigma^j + \sqrt{\frac23}\, \bar
P_{1a}^i - i \sqrt{\frac16}\, \epsilon_{ijk} \bar P_{1a}^j \sigma^k.
\end{eqnarray}
In nonrelativistic limit, the field for the $D$-wave $1^{--}$ charmonium state
can be written as~\cite{Margaryan:2013tta}
\begin{equation}
  J^{ij} = \frac12 \sqrt{\frac35} \left( \psi^i\sigma^j + \psi^j \sigma^i
\right) - \frac{1}{ \sqrt{15}} \delta^{ij} \,\vec \psi\cdot \vec\sigma,
\end{equation}
where $\psi$ annihilates the $D$-wave vector charmonium, and the spin-0 and
spin-2 states irrelevant for our study are not shown.
In order to calculate the triangle diagrams in Fig.~\ref{fig:FeynmanDiagram},
we need the Lagrangian for coupling the $D$-wave charmonia to the
$\frac12^-$-$\frac32^+$ charmed-meson pair as well as that for the E1 radiative
transitions between the charmed mesons~\cite{Guo:2013zbw}
\begin{eqnarray}
  \label{eq:lagD}
  \mathcal{L}
  = \frac{g_4}{2} \Tr \left[\left( \bar{T}_a^{j\,\dag}\, \sigma^i H^\dag_a -
 \bar{H}^\dag_a \,\sigma^i T^{j\,\dag}_a \right) J^{ij} \right]
+ \sum_a \frac{c_a}{2}\, \Tr \left[ T_a^i H_a^\dag \right] E^i
+ \text{H.c.},
\end{eqnarray}
where in the first term the Einstein summation convention is used while for the
latter we distinguish the coupling constants for different light flavors because
there is no isospin symmetry in the electromagnetic interaction.
Moreover, we parametrize the coupling of the $\X$ to the pair of vector charm
and anticharm mesons as
\begin{eqnarray}
  \label{eq:LX}
  \Lag_{X_{2}} = \frac{x_2}{ \sqrt{2} } X_2^{ij\,\dag} \left( D^{*0\,i} \bar
  D^{*0\,j} + D^{*+\,i} D^{*-\,j} \right) +  \text{H.c.}.
\end{eqnarray}

With the above preparations, we can now proceed to calculate quantitatively the
production of the $\gamma\X$ in electron--positron collisions. Although in the
heavy quark limit the production of the $D$-wave vector heavy quarkonium or the
pair of $\frac12^-$ and $\frac32^+$ heavy mesons are suppressed due to spin
symmetry~\cite{Li:2013yka}, we can expect a large spin symmetry breaking in the
charmonium mass region above 4~GeV. This may be seen from similar values of
electronic widths of the excited vector charmonia. Thus, we will assume that the
production of the $\gamma\X$ occurs through the $D$-wave charmonia or the
$D$-wave components of excited vector charmonia. Without any detailed
information about the values of the coupling constants, we can predict the
energy regions with the maximal production cross sections.
\begin{figure}[tb]
    \begin{center}
        \includegraphics[width=0.6\textwidth]{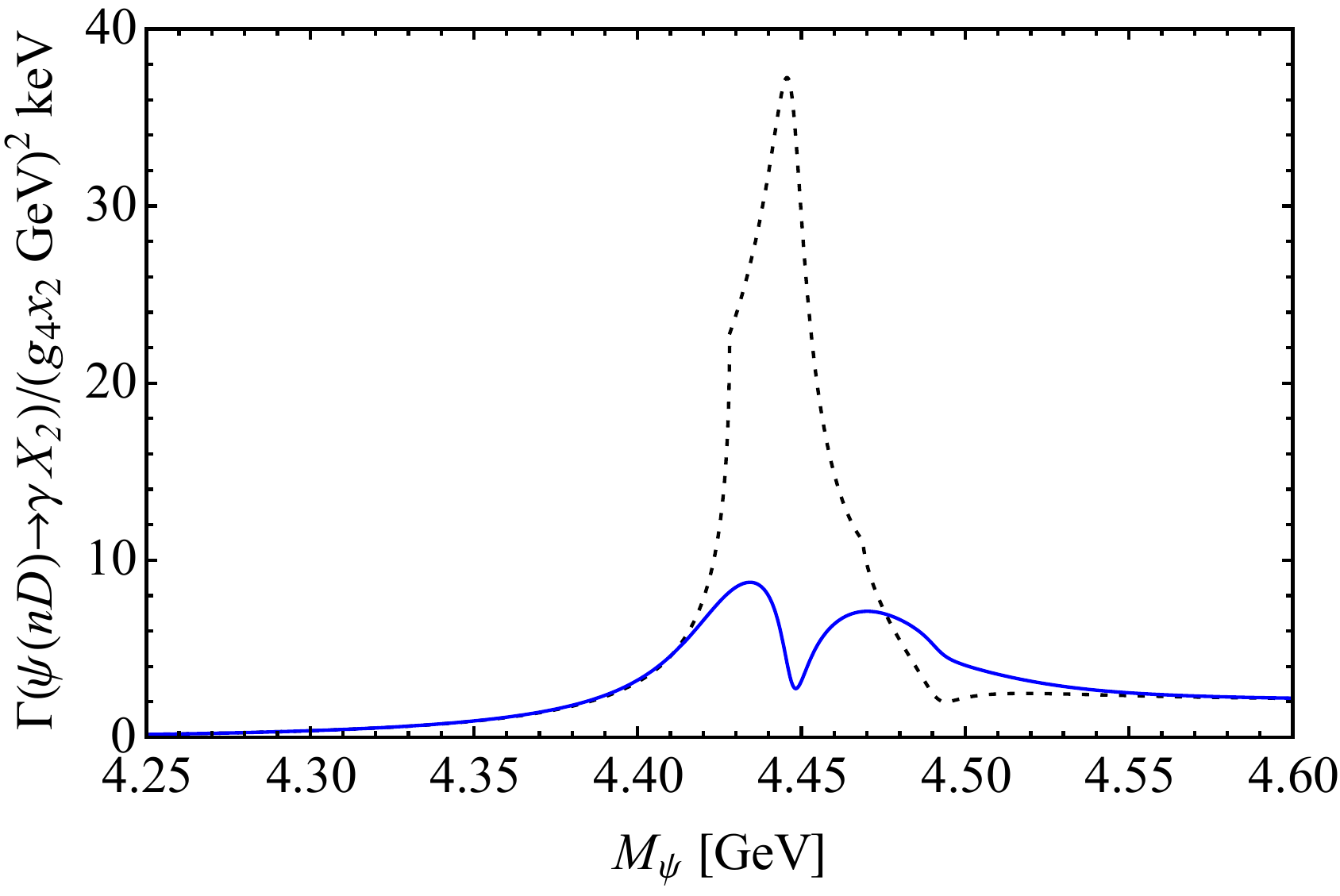}
        \caption{Dependence of the partial decay width of a $D$-wave charmonium
        into
$\gamma \X$ on the mass of the charmonium. The solid and dotted
curves are obtained with and without taking into account the widths of the
$D_1(2420)$ and $D_2(2460)$, respectively. Here, $c_u=0.4$ is used.}
\label{fig:psiDxga}
    \end{center}
\end{figure}
In Fig.~\ref{fig:psiDxga}, we show the dependence of the decay width of a
$D$-wave charmonium into the $\gamma\X$, divided by $(g_4x_2)^2$, on the mass
of the $D$-wave charmonium or the center-of-mass energy of the $e^+e^-$
collisions.
The value of the photonic coupling $c_u$ does not affect the shape of the
dependence either. Nevertheless, we took $c_u=0.4$ which is a typical value
evaluated from various quark model predictions for the decay widths $\Gamma(
D_1^0 \to \gamma
D^{(*)0})$~\cite{Fayyazuddin:1994qu,Korner:1992pz,Godfrey:2005ww}. In the
figure, the dashed curve is obtained neglecting the widths of the $D_1$ and
$D_2$ states, and the solid curve is the result of evaluating the triangle loop
integrals with constant widths for the $D_1$ and $D_2$ as
done in Ref.~\cite{Guo:2013zbw}. The maximum around 4.447~GeV and the local
minimum around 4.492~GeV of the dashed curve are due to the presence of Landau
singularities~\cite{Landau:1959fi} of triangle diagrams in the complex plane at
$(4.447\pm i\, 0.003)$~GeV (for the $D_1$ loop) and $(4.492\pm i\,0.003)$~GeV (for the $D_2$
loop), respectively (for a discussion of the Landau singularities in the
triangle diagrams of heavy quarkonium transitions, we refer to
Ref.~\cite{Guo:2014qra}).
The two cusps on both sides of the shoulders of the peak show up at the
thresholds of the $D_1\bar D^*$ and $D_2\bar D^*$.

From the figure, it is clear that the ideal energy regions for producing the
$\gamma\X$ in $e^+e^-$ collisions are around the $D_1\bar D^*$ and $D_2\bar D^*$
thresholds, i.e. between 4.4~GeV and 4.5~GeV. It is also clear
that the mass region of the $\Y$ is not good for the production of the
$\gamma\X$, contrary to the case of the $\gamma X(3872)$. In order to quantify the
relative production rate of the $\gamma\X$ with respect to the $\gamma X(3872)$,
we require the $\Y$ to couple to the $\frac12^-$-$\frac32^+$ meson pair as
follows
\begin{eqnarray}
  \label{eq:LY}
  \Lag_{Y} \al=\al \frac{y}{\sqrt{2}} Y^{i\,\dag} \left( D_{1a}^i \bar D_a - D_a
\bar D_{1a}^i \right)
  + i \frac{y^{\prime}}{\sqrt{2}} \epsilon^{ijk} Y^{i\,\dag} \left( D_{1a}^{j}
  \bar D^{*\,k}_a - D^{*\,k}_a \bar D_{1a}^{j} \right)  \nonumber\\
  \al\al+ \frac{y^{\prime\prime}}{\sqrt{2}} Y^{i\,\dag} \left( D_{2a}^{ij} \bar
  D^{*\,j}_a - D^{*\,j}_a \bar D_{2a}^{ij} \right) +  \text{H.c.},
\end{eqnarray}
where we have assumed isospin symmetry in the couplings and the flavor
index $a$ runs over up and down quarks. Notice that if the $\Y$ is a pure
$D_1\bar D$ (here and in the following the charge conjugated channels are
dropped for simplicity) molecule~\cite{Ding:2008gr,Wang:2013cya}, it would not
couple to the $D_1\bar D^*$ and $D_2\bar D^*$ as given by the $y'$ and $y''$ terms, and thus cannot decay into
the $\gamma\X$. These two terms are included to allow the decay to
occur.\footnote{The possibility of the $\Y$ to have $D_1\bar D^*$ and
$D_2\bar D^*$ components was discussed in Ref.~\cite{Cleven:2013mka}.} Because the $\X$ is
the spin partner of the $X(3872)$, for a rough estimate, we can assume that
$x_2$ takes the same value as the coupling constant of the $X(3872)$ to the
$D\bar D^{*}$. We also assume that the values of $y^{\prime}$ and
$y^{\prime\prime}$ are related to $y$ by a spin symmetry relation for $D$-wave
charmoinia. Comparing Eq.~\eqref{eq:lagD} with Eq.~\eqref{eq:LY}
one obtains $y^{\prime}=-y/2$ and $y^{\prime\prime}=\sqrt{6}y/10$.
Then, the ratio of the partial decay widths of the $\Y$ to the $\gamma\X$ and
the $\gamma X(3872)$ can be estimated parameter-free, and is
\begin{equation}
  \label{eq:wid4260}
  \frac{\Gamma( \Y \to \gamma \X)}{\Gamma( \Y \to \gamma X(3872))} \approx 10^{-2}~.
\end{equation}
In the above ratio, whether or not to take into account the finite widths of the
$P$-wave charmed mesons only results in a minor change of 2\%. It is clear that
unless the $\Y$ couples to the $D_1\bar D^*$ and/or $D_2\bar D^*$ with a
coupling much larger than that for the $D_1\bar D$, which is less possible, the
branching fraction of the $\Y \to \gamma \X$ is much smaller than that of the
$\Y \to \gamma X(3872)$. Given that the number of events for the latter process
as observed at BESIII is the order of 10~\cite{Ablikim:2013dyn}, it is unlikely
to make an observation of the $\gamma\X$ at an energy 4.26~GeV at BESIII.

To summarize, it is generally expected that the $X(3872)$ as a hadronic molecule
has a spin partner close to the $D^*\bar D^*$ threshold. In this paper, we have
investigated the production of the $\gamma\X$ in $e^+e^-$ collisions.
According to our calculation, we strongly suggest to search for the $\X$
associated with a photon in the energy region between 4.4~GeV and 4.5~GeV in
$e^+e^-$ collisions.
Besides, the width ratio of the $\Y$ decaying to $\gamma \X$ and $\gamma
X(3872)$ is quite small, at the order of $10^{-2}$. Thus observing the
$\gamma\X$ at an energy around 4.26~GeV would be unlikely in the BESIII
experiment according to the current result of $\Y\to\gamma X(3872)$.

\medskip

\section*{Acknowledgments}

This work is supported in part by the DFG and the NSFC through funds provided to
the Sino-German CRC 110 ``Symmetries and the Emergence of Structure in QCD''
(NSFC Grant No. 11261130311), by the EU Integrated Infrastructure Initiative
HadronPhysics3 (Grant No. 283286), and by NSFC (Grant No.
11165005).

\end{document}